\documentclass[a4paper,11pt]{article}
\usepackage{jcappub}

\usepackage[colorlinks=true,urlcolor=blue,linkcolor=blue]{hyperref}
\usepackage{graphicx}
\usepackage{epsfig}
\relax

\def\be{\begin{equation}}
\def\ee{\end{equation}}
\def\bs{\begin{subequations}}
\def\es{\end{subequations}}
\def\calm{{\cal M}}

\def\Lx{\Lambda}

\newcommand{\dv}{\delta{v}}
\newcommand{\dvvec}{\delta\vec{v}}
\def\Hc{{\cal H}}

\newcommand{\Vt}{\tilde V}

\newcommand{\drho}{\delta \rho}

\def\be{\begin{equation}}
\def\ee{\end{equation}}
\def\bs{\begin{subequations}}
\def\es{\end{subequations}}
\newcommand{\een}{\end{subequations}}
\newcommand{\ben}{\begin{subequations}}
\newcommand{\beq}{\begin{eqalignno}}
\newcommand{\eeq}{\end{eqalignno}}

\def \lta {\mathrel{\vcenter
     {\hbox{$<$}\nointerlineskip\hbox{$\sim$}}}}
\def \gta {\mathrel{\vcenter
     {\hbox{$>$}\nointerlineskip\hbox{$\sim$}}}}

\newcommand\fverb{\setbox\pippobox=\hbox\bgroup\verb}
\newcommand\fverbdo{\egroup\medskip\noindent%
                        \fbox{\unhbox\pippobox}\ }
\newcommand\fverbit{\egroup\item[\fbox{\unhbox\pippobox}]}
\newbox\pippobox

\def \lta {\mathrel{\vcenter
     {\hbox{$<$}\nointerlineskip\hbox{$\sim$}}}}
\def \gta {\mathrel{\vcenter
     {\hbox{$>$}\nointerlineskip\hbox{$\sim$}}}}

\title{Constraining Dark Energy through the Stability of Cosmic Structures}

\author[a]{V. Pavlidou}
\author[b,c]{N. Tetradis}
\author[d]{T.~N. Tomaras}
\affiliation[a]{Department of Physics, University of Crete, and Fo.R.T.H., 711 10 Heraklion, Greece}
\affiliation[b]{Department of Physics, University of Athens, Zographou 157 84, Greece}
\affiliation[c]{Department of Physics, CERN - Theory Division, CH-1211 Geneva 23,
Switzerland}
\affiliation[d]{Department of Physics and CCTP, University of Crete, Heraklion 711 10, Greece}
\emailAdd{pavlidou@physics.uoc.gr}
\emailAdd{ntetrad@phys.uoa.gr}
\emailAdd{tomaras@physics.uoc.gr}

\abstract{For a general dark-energy equation of state,
we estimate the maximum possible radius of massive structures that are not destabilized by the acceleration of the cosmological expansion.
A comparison with known stable structures constrains the equation of state. 
The robustness of the constraint can be enhanced through the accumulation of additional astrophysical data and a better understanding
of the dynamics of bound cosmic structures.}

\begin{document}
\begin{flushright}
CCTP-2014-24

CERN-PH-TH/2013-304
\end{flushright}

\maketitle
\flushbottom

\section{Introduction}
\setcounter{equation}{0}

The formation of structure in the Universe proceeds through the gravitational collapse of primordial
fluctuations in the energy density.
Dark energy competes against the self-gravity of fluctuations,
  and it can slow down or even prevent the collapse of overdensities
  of sufficiently low contrast with the background.
In a Universe dominated by a cosmological constant or dark
  energy, it is expected that the 
growth of structures cannot continue indefinitely (e.g.,
\cite{boundrefs0,boundrefs,boundref2,boundref3,boundref4,boundref5,pavtom}). In
later times, dark energy can become important in the dynamics of
gravitating structures \cite{bkc}
and halt the collapse on certain scales/times \cite{weinberg}.
In particular, for the outer shells of an overdense
  region the average density of which has a low enough contrast with
  the background universe, the acceleration due to the dark energy
  component can compensate for the inwards acceleration induced
  through the gravitational attraction by the inner shells.
As a result, there is a
maximum radius for an overdensity of a certain mass, beyond which
matter can only expand.
In ref. \cite{pavtom} it was suggested that $\Lambda$CDM cosmology can be tested by examining whether the mass-radius 
relation of observed structures in the Universe satisfies the constraint imposed by the above argument. 
Of the few examples that were examined, none
were  found to violate the constraint, even though certain structures are close to the boundary of the
allowed region.

The simplest way to derive the constraint of ref. \cite{pavtom} is to consider
the Schwarzschild-de Sitter metric (in a slightly different notation from the one used in \cite{pavtom} for consistency with the following sections)
with mass $\calm$ and cosmological constant $\Lx$:
\be
ds^2=-A^2(R)d t^2+B^2(R)dR^2+R^2d\Omega^2,
\label{schw} \ee
where $A^2(R)=B^{-2}(R)=1-\calm/(8\pi M^2R)-\Lx\, R^2/3$.
$d\Omega^2$ is the metric on a two-sphere, while Newton's constant is $G=1/16\pi M^2$. The speed of light is set to $c=1$ throughout. 
The metric (\ref{schw}) describes the exterior of a 
compact spherically symmetric object of mass $\calm$ in the presence of the cosmological constant.
An observer located at a constant value
of $R$  in the system of coordinates (\ref{schw}) has four-velocity $u^\alpha=(1/A,0,0,0)$ and four-acceleration
\be
u^\alpha_{~;\beta}u^\beta=\left( 0, \frac{G \calm}{R^2}-\frac{\Lx  R}{3},0,0\right),
\label{accschw} \ee
while, by definition, the four-acceleration of a freely falling one vanishes.
Thus, the outward acceleration of the observer at constant $R < (3 G \calm/\Lx)^{1/3}$ is due to  
an outward external force that counter-balances the gravitational attraction of the mass. On the other hand, an observer
at constant $R > (3 G \calm/\Lx)^{1/3}$ must be subject to an inward
force that balances the 
influence of the cosmological constant. It is obvious then that a
freely falling observer initially located at rest at 
$R < (3 G \calm/\Lx)^{1/3}$ will move towards the central mass, while
a freely falling observer initially located at 
$R > (3 G \calm/\Lx)^{1/3}$ will move away from it. The value 
\be
R_c = \left( \frac{3 G \calm}{\Lx} \right)^{1/3}
\label{rc} \ee
determines the outer radius at which the gravitational attraction of the central mass exactly balances the effect of the cosmological
constant. The radius of the bound object cannot exceed this value, as its outer layers would be destabilized by the expansion.

For a time-dependent configuration, the value (\ref{rc}) corresponds
to the maximum possible value of the turnaround radius of an
overdensity. The turnaround radius is the one reached when its
outer layers stop expanding and start collapsing
under  the overdensity's self-gravity. This bound on the turnaround radius 
depends only on the total mass it encloses and the value of the cosmological constant \cite{pavtom}. 
In Section \ref{lambda}
we review the evolution of dark matter structures much earlier than
virialization, starting from the Lemaitre-Tolman-Bondi metric (LTB) \cite{ltb,ltb2,ltb3} for a pressureless fluid in a 
spacetime with a cosmological constant. In this way we verify that the
maximum turnaround radius in such a setting is given by eq. (\ref{rc}), with $\calm$
corresponding to the enclosed mass.
We then, in Section \ref{dark},
  generalize the discussion to the case of dark energy with a general
  equation of state. Exact generalizations of the LTB metric in the presence of fluids with pressure  
are possible, but they involve a large number of functions to be determined through the Einstein equations \cite{hellaby}.
Allowing for a generic equation of state for the background makes an
analytical treatment technically difficult.  
However, as we shall show in section \ref{dark}, an explicit solution can be derived using a perturbative approach that
captures all the important features of the problem. Moreover, this approach makes it possible to relax the
assumption of spherical symmetry. 

The generalization of eq. (\ref{rc}) that we shall derive can be used 
in order to obtain a constraint on the equation of state 
$w=p/\rho$. This can be achieved by checking whether the radius of a massive observed structure is indeed smaller than
the expected turnaround radius. In section \ref{observations} 
we shall compare with the data summarized in ref. \cite{pavtom}, as well as with additional data from the literature. 
We shall discuss the constraint imposed on $w$ and the prospects for making it more robust.

\section{Turnaround radius in a Universe with a cosmological constant} \label{lambda}
\setcounter{equation}{0}

We consider a spherically symmetric configuration of a pressureless fluid in a background with a cosmological 
constant. 
For the metric we use the ansatz
\begin{equation}
ds^{2}=-dt^2+\frac{R'^2(t,r)}{1+f(r)}dr^2+R^2(t,r)d\Omega^2,
\label{metrictbl}
\end{equation}
(LTB metric) where $d\Omega^2$ is the metric on a two-sphere, 
$f(r)$ is an arbitrary function, and 
the prime denotes differentiation with respect to $r$.
The energy momentum tensor has the form
\begin{equation}
T^A_{~B}={\rm diag} \left(-\rho(t,r)-2M^2\Lx,\, -2M^2\Lx,\, -2M^2\Lx,\, -2M^2\Lx  \right).
\label{enmomtb} \end{equation}
The fluid consists of successive shells marked by $r$, whose  
local density is time-dependent. 
The function $R(t,r)$ describes the location of the shell $r$
at time $t$. Through an appropriate rescaling it can be chosen to satisfy
$R(t_0,r)=r$ at some initial time $t_0$.

From the Einstein equations for the above $T^A_{\;\,B}$ one can derive
\be
\frac{1}{R^2R'}\left[R' \left(\dot{R}^2-f \right)+R \left(2\dot{R}\dot{R}'-f' \right) \right]=
\frac{1}{2M^2}\rho+\Lx, 
\label{eins1} \ee
where the dot denotes differentiation with respect to $t$, together with 
the conservation of the energy-momentum tensor
\be
R^2 R'\dot{\rho}+2 RR'\dot{R} \rho+R^2\dot{R}'\rho=
\frac{\partial}{\partial t}\left(R^2 R'\rho \right)=0.
\label{constmn} \ee
Integration of the latter equation leads to 
\be
\rho(t,r)=\frac{\calm'(r)}{4\pi R^2(t,r)R'}
\label{massr} \ee
with an arbitrary function $\calm(r)$, and upon 
substitution in eq. (\ref{eins1}) one obtains 
\be
\frac{\partial}{\partial r}\left(R\dot{R}^2 - Rf \right)=\frac{1}{2M^2} \frac{\partial}{\partial r}
\left( \frac{\calm}{4\pi} +\frac{2}{3}M^2\Lx R^3 \right).
\label{soll} \ee
This is integrated trivially and leads to the effective Friedmann equation
\be
\frac{\dot{R}^2(t,r)}{R^2}=\frac{1}{6M^2} \frac{\calm (r)}{\frac{4\pi R^3}{3}} +\frac{\Lx}{3} +\frac{f(r)}{R^2},
\label{tb1} \ee
with 
$1/G=16 \pi M^2 $.
The  function $\calm(r)$ of the fluid can be chosen 
arbitrarily. It incorporates the
contributions of all shells up to $r$ and determines the energy density
through eq. (\ref{massr}). Because of energy conservation, $\calm(r)$  
is independent of $t$, while $\rho$ and $R$ depend on both $t$ and $r$.
The function $f(r)$ acts as an effective curvature term. It can also be considered as part of the initial 
velocity of the fluid.
It is straightforward to check that eqs. (\ref{massr}), (\ref{tb1}) imply the complete set of Einstein field equations.

The two arbitrary functions $\calm(r)$, $f(r)$ can be used in order to model a large variety of expanding or collapsing
cosmological configurations. We are interested in the evolution of an initially expanding overdensity. The function 
$\calm(r)$ can be written as $\calm(r)=\calm_0(r)+\delta\calm(r)$, where $\calm_0(r)=4\pi r^3 \rho(t_0)/3$  is the contribution from the
homogeneous part of the energy density at some initial time $t_0$, while $\delta\calm(r)>0$ approaches a 
constant value for $r\to\infty$. The function $f(r)$ must be taken negative for the solution of eq. (\ref{tb1}) to describe the 
turnaround and eventual collapse of the shells of the overdense region. 
A very intuitive picture is obtained if we assume that at the 
initial time $t_0$ all shells have the same expansion rate, determined by the average density: $H^2_0=\rho(t_0)/(6M^2)+\Lx/3$.
Then, from eq. (\ref{tb1}) we obtain $f(r)=-\delta\calm(r)/(8\pi M^2 r)$. With this assumption, the evolution of each shell and its
possible turnaround is determined by the mass excess it encloses.

For $f(r)<0$, a shell marked by $r$ will turn around and start collapsing 
if at some time $t>t_0$ there is a solution to the condition $\dot{R}(t,r)=0$. As can be derived from eq. (\ref{tb1}), 
the maximum turnaround radius corresponds to
the smallest value of $\kappa=8\pi M^2 |f(r)|/\calm(r)$ for which the equation $\omega R^3 -\kappa R+1=0$, with
$\omega=8\pi M^2\Lx/(3\calm(r))$, has a real positive solution. As $|f(r)|/\calm(r)$ is a decreasing function of $r$ 
for the physical situation we are considering, the maximum turnaround radius is obtained for a certain shell $r_c$, with all 
shells $r>r_c$ never collapsing. The minimal value of $\kappa$ is $\kappa_{min}=3 \omega^{1/3}/2^{2/3}$, and
the corresponding physical distance of the shell $r_c$ 
\be
R_c\equiv R(t_c,r_c)=\left( \frac{3G\calm(r_c)}{\Lx}  \right)^{1/3}
\label{accelr} \ee
at the turnaround time $t_c$. The detailed form of $f(r)$ does not affect this relation.

From eq. (\ref{tb1}) we can easily compute 
\be
\ddot{R}(t,r)=-\frac{1}{16\pi M^2}\frac{\calm(r)}{R^2}+\frac{1}{3}\Lx R.
\label{accel} \ee
The acceleration of the shell $r_c$ vanishes exactly at its turnaround time. The shells $r<r_c$ reach the turnaround point 
while their expansion has nonzero deceleration. 
The similarity with the discussion in the introduction indicates that eq. (\ref{accelr}) is a generalization 
of eq. (\ref{rc}) for a dynamical configuration with a continuous mass distribution in space. 
The distance $R_c$ corresponds to the maximum radius of the outmost shell that can decouple from the 
general expansion and collapse towards the center of an overdensity. 

If the integrated mass is dominated by the contribution from the
central region, where the energy density is very large, $\calm$ grows very slowly for large $r$, so that the exact determination of the
value $r_c$ is not necessary. In this case, eq. (\ref{accelr}), with approximately constant $\calm$,
determines the physical distance $R_c$ beyond which matter can never collapse, as it continuously accelerates outwards.

\section{Turnaround radius in a Universe with dark energy} \label{dark}
\setcounter{equation}{0}

In this section we generalize our previous discussion to the case of dark energy. As we mentioned in
the introduction, an exact solution of the Einstein equations is not feasible in this case. For this reason we employ a
perturbative approach, inspired by the theory of cosmological perturbations in the Newtonian gauge. The 
details of the scheme can be found in the appendix A of ref. \cite{np}.

We assume that the energy density of the Universe 
receives significant contributions from two components: 
a) the nonrelativistic matter, consisting of standard baryonic matter and cold dark matter; and  
b) a homogeneous component characterized as dark energy. 
%We employ the reduced Planck mass $M=(16\pi G)^{-1/2}$. 
For the metric we consider an ansatz of the form
\be
ds^2=a^2(\tau)\left[
-\left(1+2\Phi(\tau,\vec{x}) \right)d\tau^2
+\left(1-2\Phi(\tau,\vec{x}) \right) d\vec{x}\, d\vec{x} \right].
\label{metric} \ee
We assume that the Newtonian potential $\Phi$ is weak, $\Phi \ll 1$ and that the
dark energy is homogeneous with density $\rho_{E}(\tau)$.
 The energy density of matter is decomposed as
\be
\rho_M(\tau,\vec{x})=\rho(\tau)+\drho(\tau,\vec{x}).
\label{decomrho} \ee
We allow for significant
density fluctuations, even though our analysis loses accuracy when they
are much larger than the background density.  
We would like to take into account the effect of the local velocity field
$\dvvec$
when it becomes significant because of large field gradients. 
We expect that the velocity field is driven by the spatial
derivatives of the potential $\Phi$. The potential, in its turn, is sourced by the 
inhomogeneities in the matter distribution.
For subhorizon perturbations with momenta $k\gg \Hc=\dot{a}/a$ (in the present section the dot denotes differentiation
with respect to $\tau$), 
we expect 
$|\dvvec| \sim (k/\Hc) \Phi \sim (\Hc/k)(\drho/{\rho})$. 
Consistently with this expectation, we assume the  
hierarchy of scales: $\Phi \ll |\dvvec | \ll \drho/{\rho}  \lta 1$. 
As we are dealing with subhorizon perturbations, it is also consistent to make
the additional assumption that the spatial derivatives of $\Phi$
dominate over its time derivative.

The energy-momentum tensor of matter is 
$ \left( T_{M} \right)^{\mu}_{~\nu}= \rho\, \Vt^\mu \Vt_\nu$. 
We define the peculiar velocity through $\Vt^i=\dv^i(\tau,\vec{x})/a(t)$.
Thus, at linear order, the energy-momentum tensor of matter becomes
\begin{eqnarray}
\left( T_{M} \right)^0_{~0}&=&-{\rho}-\drho
\nonumber \\
\left( T_{M} \right)^0_{~i}&=&{\rho}\,\dv_i
\nonumber \\
\left( T_{M} \right)^i_{~j}&=&0,
\label{tdm} \end{eqnarray}
with $\dv_i=\dv^i$, while the energy-momentum tensor of dark energy is 
$ \left( T_{E} \right)^{\mu}_{~\nu}= {\rm diag}( -\rho_E,  p_E, p_E, p_E)$. 

With the above assumptions, one can derive the equations 
that describe the evolution of the Universe. 
The homogeneous background satisfies the well known Friedmann and energy conservation equations
\begin{eqnarray}
\Hc^2\equiv \left( \frac{\dot{a}}{a} \right)^2&=&
\frac{1}{6 M^2} a^2 \left( \rho+\rho_E \right),
\label{hubble} \\
\dot{\rho}+3\mathcal{H}\rho
&=&\dot{\rho}_E+3\mathcal{H}\left( \rho_E+p_E\right)=0.
\label{25c} \end{eqnarray}
The linear order Einstein equation for $\Phi$ is
\be
\nabla^2 \Phi -3 \Hc\dot{\Phi}-3\Hc^2\Phi=\frac{1}{4 M^2} a^2\drho\,,
\label{gravpot}
\ee
which, given our assumption about the relative size of spatial and time derivatives, reduces to 
 the Poisson equation:
\be
\nabla^2\Phi=\frac{1}{4 M^2} a^2 \drho.
\label{gravpois} \ee
From the conservation of the energy-momentum tensor we obtain the linearized continuity and Euler equations 
\begin{eqnarray}
{\delta\dot{\rho}}+3\mathcal{H}\delta\rho+\rho \vec{\nabla}\delta{\vec{v}_i}&=&0
\label{a}\\
 \delta \dot{\vec{v}}+\mathcal{H} \delta \vec{v}
&=&-\vec{\nabla}\Phi.
\label{b}  \end{eqnarray}

The peculiar velocity $\Vt^i=dx^i/dt=\delta v^i/a$, with $dt=a(\tau)d\tau$, has been defined in terms of the comoving coordinates $x^i$. 
Therefore, the velocity corresponding to physical distances $R^i=a(\tau)x^i$ is $V^i=dR^i/dt=\Hc x^i+\delta v^i$ and the corresponding acceleration has the form:
\be
\frac{d^2 \vec{R}}{dt^2}=\left(\frac{\ddot{a}}{a^2}-\frac{\dot{a}^2}{a^3}\right) \vec{x}
+\frac{1}{a}\left( \delta \dot{\vec{v}}+\mathcal{H} \delta \vec{v} \right)
=-\frac{1}{12 M^2}(\rho+\rho_E+3p_E)\vec{R}   -\frac{1}{a}\vec{\nabla}\Phi.
\label{acc} \ee
Eqs. (\ref{gravpois}), (\ref{acc}) are the generalization of the
linearized form of eq. (\ref{accel}) for a background with an arbitrary equation of
state and a generic density fluctuation not necessarily spherically symmetric. 

For a spherically symmetric configuration, in particular, one obtains
\be
\frac{d^2 R}{dt^2}=-\frac{1}{12 M^2}(\rho_E+3p_E) R   
-\frac{1}{16\pi M^2} \frac{\calm(r)}{R^2},
\label{accsph} \ee
where in the integrated mass $\calm$ we have incorporated the contribution of the homogeneous matter density $\rho$. 
This equation  extends eq. (\ref{accel}) to the case of dark energy. 
If $\calm$ is dominated by the contribution from the central region, we can approximate it as constant
for sufficiently large $r$. In this case, the physical distance $R_c$ at which the acceleration vanishes determines the 
boundary of the region within which matter can decouple from the expansion and collapse. We find
\be
R_c=\left( -\frac{3\calm}{4(1+3w)\pi \rho_E} \right)^{1/3},
\label{accelde} \ee
where we have assumed that $w<-1/3$. For the special case of a cosmological constant, i.e. for $w=-1$ and $\rho_E=2 M^2 \Lx$, one recovers eq. (\ref{accelr}).

\section{Observational constraint on the equation of state} \label{observations}
\setcounter{equation}{0}

\begin{figure}[t]
\begin{center}
 \epsfig{file=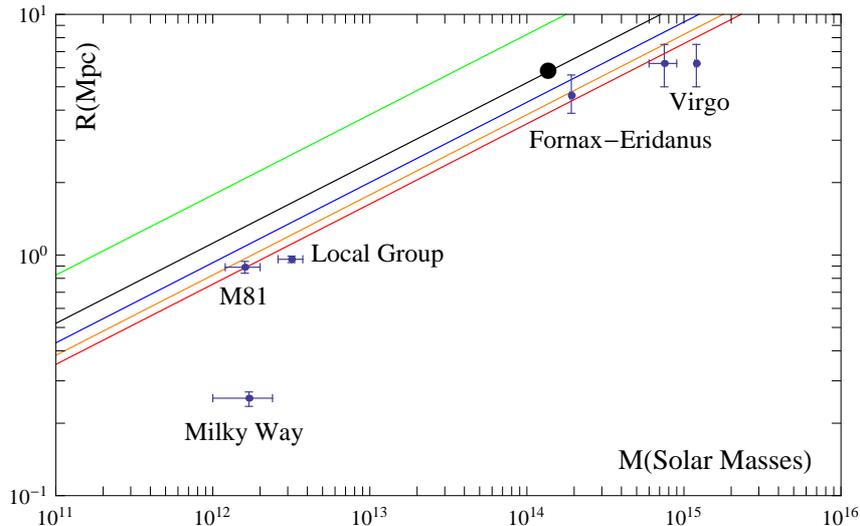,width=0.7\linewidth,clip=}
\end{center}
\caption{ The bound on the mass-to-radius relation of stable
  structures for various values of the parameter $w$ of the
  dark-energy equation of state. The successive lines, starting from the top, correspond to $w=-0.5, -1, -1.5, -2$ and $-2.5$.
Masses and radii of known structures are also depicted (see text for references). }
 \label{fig1}
 \end{figure}

We can use eq. (\ref{accelde}) to constrain the parameter $w$ of the
equation of state. This can be achieved by comparing with masses and radii of known cosmic structures.
For a given mass, eq. (\ref{accelde}) determines the upper limit on the possible radius of an object that has stopped expanding. We shall focus on structures in our vicinity, characterized by low redshift. In this way, a constraint can be placed on the
  equation of state of dark energy using current-cosmic-epoch
  data. Such a constraint is largely orthogonal to the constraints from the cosmic
  microwave background that use data from the epoch of last
  scattering, involving very different uncertainties. 
We use some of the data summarized in ref. \cite{pavtom}, as well as additional data from the literature. 

The turnaround radius of a given structure can be estimated from the average radius $R_0$ of the zero-velocity surface, which
delineates the region of space that does not follow the general cosmological expansion. The average velocity of
objects in the neighborhood of a massive structure deviates from the linear Hubble relation and eventually crosses zero at $R_0$. 
Estimates of  the value of $R_0$ can be obtained through observations of nearby flows.
The mass of a given structure is often deduced from the radius of the zero-velocity surface by identifying it with the 
turnaround radius and using some incarnation of the spherical
  collapse model to relate this radius to the enclosed mass.
Values obtained in this way 
enforce by construction the spherical collapse
predictions for $w=-1$, and, therefore, 
cannot be used for constraining the equation of state through our procedure. 
In order to constrain the equation of state, one must employ mass estimates obtained independently through alternative methods. 

The virial relation is a commonly used mass estimator for systems that are dynamically relaxed. It is applicable in the central regions of 
galaxies or clusters, where astrophysical objects have decoupled from the general expansion and the average matter density is much larger than
the dark energy density.  Other similar methods are also available, based on the modeling of orbital motions. 
Additional input on the distribution of 
matter, for example from simulations of structure formation, can
improve the precision of such estimates.
Other methods, that have been used as mass estimators, are the X-ray emission profile of the hot intra-cluster gas which 
depends on the gravitational potential of clusters, and strong or weak gravitational lensing. However, the quoted errors in the values
deduced through these methods are rather large. The weak point of these mass estimators for our purposes 
is that they do not take into account the matter distributed between the virial and turnaround radii.

The data that provide interesting constraints on the equation of state through our
approach have been obtained primarily through the analysis of virial or orbital motions. 
We provide here a brief summary of the data that we use:
\begin{itemize}
\item 
For the radius of the Milky Way 
we use $R=254^{+16}_{-19}$ kpc, corresponding to the distance of the Leo I dwarf
spheroidal galaxy, which we assume to be bound to the Milky Way \cite{milkyway,milkyway2}.
We also assume a mass range $\calm=(1-2)\times 10^{12} M_\odot$, which can fit most observational data, 
as summarized in ref. \cite{milkyway,milkyway2}. 
\item 
For the radius of the Local Group we take the radius of the zero-velocity surface
$R=0.96\pm 0.03$ Mpc, as given in ref. \cite{localm81}. The mass estimate of \cite{localm81} is deduced from this 
radius and cannot be used as a constraint within our approach. For this reason, we employ the kinematic estimate  
 $\calm=(3.17\pm0.57) \times 10^{12} M_\odot$ of ref. \cite{local}. 
\item
For the M81/M82 Group we take the radius of the zero-velocity surface  $R=0.89\pm 0.05$ Mpc 
from ref. \cite{localm81} and the mass estimate
$\calm=(1.2-2)  \times 10^{12} M_\odot$ from ref. \cite{m81}, which is derived via the virial theorem and from orbital motions.
\item
For the Virgo cluster, ref. \cite{virgor0} gives an estimate for the radius of the zero-velocity surface in the range
$R=5-7.5$ Mpc. The virial mass estimate $\calm=(7.5\pm 1.5)\times 10^{14} M_\odot$ is given in ref. 
 \cite{virgo2}. It is consistent with the value 
$\calm=7 \times 10^{14} M_\odot$ derived in ref. \cite{virgo3} from orbital motions.
Note that the use of the Tolman-Bondi model for modelling the cluster in  
ref. \cite{virgo} results in a much larger value $\calm=1.2\times
10^{15} M_\odot$ within 2.2 Mpc radius (not a turnaround radius).
\item
In ref. \cite{fornax},  the radius $R=4.6$ Mpc of the zero-velocity surface for the Fornax-Eridanus complex is given, with a confidence interval of 
$3.88-5.60$ Mpc. 
The mass deduced from this radius cannot be used for our purposes. However, the estimate $\calm=1.92 \times 10^{14} M_\odot$ for the total
mass of all virialized groups bound to the Fornax cluster is also given in ref. \cite{fornax}, following the analysis of ref. \cite{makarov}. 
This value is compatible with the estimate from the zero-velocity radius.

\end{itemize}

We plot these data in Fig. \ref{fig1}. For the Virgo cluster we depict points corresponding to the mass estimates of both refs. \cite{virgo2} and 
\cite{virgo}. 
In the same figure we also plot the lines corresponding to the relation (\ref{accelde}) for a Hubble parameter
$H_0=67.3$ km/s/Mpc, and matter and dark energy density parameters $\Omega_m=0.315$ and $\Omega_E=0.685$, respectively.
The successive lines, starting from the top, correspond to $w=-0.5, -1, -1.5, -2$ and $-2.5$.
The part of the parameter space above each of the lines in Fig. \ref{fig1} corresponds to the range for which no stable structures should exist for the corresponding $w$. Very negative values of $w$ result in extreme acceleration, which is incompatible with some of the observed 
structures. 

In the low mass range, 
the M81/M82 Group limits $w$ to values $w\gta -2.5$, while 
the Local Group lies close to the $w=-2.5$ line. The Milky Way is well within the allowed region. This indicates that matter in 
more compact objects, such as individual galaxies, has collapsed well below the maximum turnaround radius. The corresponding data
do not provide a significant constraint on the equation of state of dark matter. 
On the other hand, groups of galaxies are more extensive structures, from which 
meaningful bounds for $w$ can be obtained.
In the higher mass range, the various estimates for the Virgo cluster are consistent with $w\gta -2.5$, even though the limited
precision of the data indicates that the constraint could be tightened.
The Fornax-Eridanus complex requires $w\gta -2$, even though the precision is limited again. 

A very interesting possibility is discussed in ref. \cite{eridanus}, where it is suggested
that galaxy groups in the 
Eridanus constellation are in the process of forming a supergroup. 
The analysis indicates that there is 
a 56\% probability that the Eridanus, NGC 1407 and NGC 1332 groups form a supergroup which is bound and eventually will merge
with the Fornax cluster. The mass of the supergroup-Fornax structure is $\calm = 13.7 \times 10^{13} M_\odot$ and its projected radius
$R=5.9$ Mpc. Even though the structure is at an early evolutionary stage, it appears to have decoupled from the 
general expansion.  The data point for this structure is depicted by a dark circle in Fig. \ref{fig1}.
It falls on the $w=-1$ line, leading to a very strong constraint on the equation of state. 
It must be noted, however, that this large structure is still evolving and the analysis providing its mass and radius is subject to 
large uncertainties. In particular, the effect of the dark energy on the dynamics of the structure is not taken into account in ref. \cite{fornax}, even
though the average energy density of matter is comparable to that of dark energy at these length scales.
The absence of error estimates for the mass and radius given in ref. \cite{eridanus} makes it clear that a more thorough analysis is required before
definite conclusions can be drawn. On the other hand, it seems plausible that the best bound on the equation of state can be 
obtained from structures that are at the early stages of gravitational collapse and have a significant amount of matter distributed 
at distances comparable to the turnaround radius. Such structures are still evolving and very good modeling is required in order to
determine their mass and radius with precision. 

That the mass estimates used in Fig. \ref{fig1} do not correspond to
the radii to which they are matched (i.e., that the masses are estimated
out to a different radius than the one depicted in the figure) is a
general caveat of our conclusions regarding constraints on $w$. If the
additional mass between (a) the outermost radius corresponding to the mass estimates
and (b) the turnaround radius is {\em not appreciable} (which is more likely
the more relaxed the structure is), then our constraints are valid. If
on the other hand the additional mass above {\em is a significant fraction}
of the estimated mass, then the datapoints in Fig. \ref{fig1} have to
be transported to the right and the corresponding constraints
become weaker.

Deviation from spherical symmetry is another effect that can alter
the strength of the constraints discussed above. In fact,
it was shown in ref. \cite{boundref3} that, in the limit of a highly
flattened fluctuation, turnaround and collapse occur regardless of
the value of the cosmological constant. However, this is not
predictive of the behavior of a region within a spherical boundary
around a collapsed structure (which is the observable we are
after). Ref. \cite{boundref2} provides data from cosmological simulations
for the size of the ``sphere of gravitational influence'' (in essence equivalent to
our maximum turnaround radius), which by construction include the
effect of nearby structures and deviations from spherical
symmetry. It is found that, for realistic structures with mass between
$10^{14}$ and $10^{16} {\, \rm M\odot}$, the ``sphere of influence'' is no more
than 30\% larger than the result assuming exact spherical
symmetry. Ideally, the problem can be addressed by targeted analysis of
(already existing) cosmological simulations, matching the mass ranges of interest as well as the way the
turnaround radius is defined in simulations and observations of asymmetric structures.

Another important observation is that the method can be applied also to the case of a varying equation of state for
dark energy. The perturbative equations employed in section \ref{dark} have been generalized in ref. \cite{brouz} for 
a dark energy sector modelled as an evolving scalar field with a nonzero potential. They can be recast as 
evolution equations for a dark energy fluid with a variable equation of state $w(a)$, with $a$ the scale factor.
It is straightforward to see that  the maximum turnaround radius at a given time is given by eq. (\ref{accelde}), with $w=w(a)$. 
It is, therefore, possible to employ structures at high redshifts as well, in order to constrain the time evolution of the 
equation of state of dark matter. 

It is clear that more astrophysical data are needed for the bound on $w$ to become robust.
The standard approach on constraining the cosmological scenario through the properties of large cosmic structures focuses on
determining the number of massive objects at various redshifts and examining whether it is consistent with the theory of
structure formation. A precise comparison of theory and observations
requires a good understanding of the process of structure formation, especially during the
stages that cosmological perturbation theory is not applicable. Numerical simulations is the only reliable, but technically difficult and time consuming,
method for making theoretical predictions.
On the other hand, the argument of ref. \cite{pavtom} that we generalized for the dark-energy scenario 
is characterized by its simplicity and generality.
It does not depend on the details of the cosmological model or the assumed primordial spectrum. It is essentially Newtonian, so
that it applies even if the dark energy is viewed as a parametrization of deviations from General Relativity at very large scales.  
In addition, the observables involved (total enclosed mass and enclosing radius)
are in general more easily measured than, for
example, the density profile of structures at similar scales \cite{ostriker, prada, kravtsov}.

A detailed quantitative assessment of the power of the approach
presented here to yield constraints on $w$ is essential for its future applications. This process will involve both numerical
simulations (to assess the effect of deviations from sphericity and
that of neighboring structures, as discussed above), as well as a
systematic study of observational techniques for measuring the mass
and turnaround radius of structures at different mass ranges and redshifts. 

%\section{Conclusions}
%\setcounter{equation}{0}

\acknowledgments
V.P. would like to thank Brian Fields for useful discussions. V.P. acknowledges support by European Union's
Seventh Framework Programme under grant agreements
PCIG10-GA-2011-304001 and PIRSES-GA-2012-316788, and by the European Union
(European Social Fund, ESF) and Greek national funds through the
Operational Program ``Education and  Lifelong Learning'' of the
National Strategic Reference Framework (NSRF) under the  ``ARISTEIA''
Action.  
%We would like to thank ... for useful discussions.
The work of N.T. has been supported in part 
%by the ITN network ``UNILHC'' (PITN-GA-2009-237920) and 
by the European Commission under the ERC Advanced Grant BSMOXFORD 228169.
It has also been co-financed by the European Union (European Social Fund – ESF) and Greek national 
funds through the Operational Program ``Education and Lifelong Learning" of the National Strategic Reference 
Framework (NSRF) - Research Funding Program: ``THALIS". Investing in the society of knowledge through the 
European Social Fund''. The research of T.N.T. 
is implemented under the ``ARISTEIA II" Action of the Operational Program ``Education and Lifelong Learning" and is co-funded by the European Social Fund (ESF) and National Resources. He is also supported by the EU program ``Thales" ESF/NSRF 2007-2013.

%\newpage


\begin{thebibliography}{999}

\bibitem{boundrefs0} 
J.~C.~Jackson, 
Mon.\ Not.\ Roy.\ Astron.\ Soc.\  {\bf 148} (1970) 249.

\bibitem{boundrefs} 
O.~Lahav, P.~B.~Lilje, J.~R.~Primack and M.~J.~Rees,
Mon.\ Not.\ Roy.\ Astron.\ Soc.\  {\bf 251} (1991) 128.

\bibitem{boundref2} M.~T.~Busha {\it et al.}, 
 Astrophys. J. {\bf 596} (2003) 713.

\bibitem{boundref3} 
 J.~D.~Barrow  and P.~Saich, 
Mon.\ Not.\ Roy.\ Astron.\ Soc.\  {\bf 262} (1993) 717.

\bibitem{boundref4}
V.~R.~Eke, S.~Cole and C.~S.~Frenk, 
Mon.\ Not.\ Roy.\ Astron.\ Soc.\  {\bf 282} (1996) 263.

\bibitem{boundref5} 
V.~Pavlidou and B.~D.~Fields,
Phys.\ Rev.\ D {\bf 71} (2005) 043510.  

\bibitem{pavtom}
  V.~Pavlidou and T.~N.~Tomaras,
  %``Where the world stands still: turnaround as a strong test of \Lambda CDM cosmology,''  
arXiv:1310.1920 [astro-ph.CO].  
%%CITATION = ARXIV:1310.1920;%%

\bibitem{bkc} 
G.~S.~Bisnovatyi-Kogan and A.~D.~Chernin,  
Astrophys. \& Space Sc. {\bf 338} (2012) 337. 

\bibitem{weinberg}   
S.~Weinberg,
  %``Anthropic Bound on the Cosmological Constant,''
  Phys.\ Rev.\ Lett.\  {\bf 59} (1987) 2607.
  %%CITATION = PRLTA,59,2607;%%

\bibitem{ltb}
G.~Lemaitre, 
Gen. \ Rel. \ Grav. \ {\bf 29} (1997) 641
\bibitem{ltb2}
R.~C.~Tolman,
%``Effect Of Imhomogeneity On Cosmological Models,''
Proc.\ Nat.\ Acad.\ Sci.\  {\bf 20} (1934) 169
%%CITATION = PNASA,20,169;%%

\bibitem{ltb3}H.~Bondi,
%``Spherically Symmetrical Models In General Relativity,''
Mon.\ Not.\ Roy.\ Astron.\ Soc.\  {\bf 107} (1947) 410.
%%CITATION = MNRAA,107,410;%%

\bibitem{hellaby}
  A.~A.~H.~Alfedeel and C.~Hellaby,
  %``The Lemaitre Model and the Generalisation of the Cosmic Mass,''  
Gen.\ Rel.\ Grav.\  {\bf 42} (2010) 1935  [arXiv:0906.2343 [gr-qc]].  
%%CITATION = ARXIV:0906.2343;%%

\bibitem{np}
  F.~Saracco, M.~Pietroni, N.~Tetradis, V.~Pettorino and G.~Robbers,
  %``Non-linear Matter Spectra in Coupled Quintessence,''  
Phys.\ Rev.\ D {\bf 82} (2010) 023528  [arXiv:0911.5396 [astro-ph.CO]].  
%%CITATION = ARXIV:0911.5396;%%

\bibitem{planck}
  P.~A.~R.~Ade {\it et al.}  [Planck Collaboration],
  %``Planck 2013 results. XVI. Cosmological parameters,''
  arXiv:1303.5076 [astro-ph.CO].
  %%CITATION = ARXIV:1303.5076;%%

\bibitem{milkyway} 
  M.~Boylan-Kolchin, J.~S.~Bullock, S.~T.~Sohn, G.~Besla and R.~P.~van der Marel,
  %``The Space Motion of Leo I: The Mass of the Milky Way's Dark Matter Halo,''
  Astrophys.\  J.\  {\bf 768} (2013) 140
  [arXiv:1210.6046 [astro-ph.CO]];
  %%CITATION = ARXIV:1210.6046;%%

\bibitem{milkyway2}
  M.~Bellazzini, N.~Gennari, F.~R.~Ferraro and A.~Sollima,
  %``The Distance to the Leo I dwarf spheroidal galaxy from the red giant branch tip,''
  %Submitted to: Mon.Not.Roy.Astron.Soc.
 Mon.\ Not.\ Roy.\ Astron.\ Soc.\  {\bf 354} (2004) 708
  [astro-ph/0407444].
  %%CITATION = ASTRO-PH/0407444;%%

\bibitem{localm81}
  I.~D.~Karachentsev and O.~G.~Kashibadze,
  %``Total masses of the local group and m81 group derived from the local hubble flow,''
 Astrophysics {\bf 49 } (2006) 3
  [astro-ph/0509207].
  %%CITATION = ASTRO-PH/0509207;%%

\bibitem{local}
  R.~P.~van der Marel, M.~Fardal, G.~Besla, {\it et al.}, 
%R.~L.~Beaton, S.~T.~Sohn, J.~Anderson, T.~Brown and P.~Guhathakurta,
  %``The M31 Velocity Vector. II. Radial Orbit Towards the Milky Way and Implied Local Group Mass,''
 Astrophys. J. {\bf 753} (2012) 8
  [arXiv:1205.6864 [astro-ph.GA]].
  %%CITATION = ARXIV:1205.6864;%%

\bibitem{m81}
I.~D.~Karachentsev, A.~E.~Dolphin, D.~Geisler,  {\it et al.}, 
%Grebel, E. K., Guhathakurta, P., Hodge, P. W., Karachentseva, V. E., Sarajedini, A., Seitzer, P. & Sharina, M. E. 
 Astronomy and Astrophysics {\bf 383} (2002) 125.

\bibitem{virgor0}
  I.~D.~Karachentsev and O.~G.~Nasonova,
  %``The observed infall of galaxies towards the Virgo cluster,''
Mon.\ Not.\ Roy.\ Astron.\ Soc.\  {\bf 405} (2010) 1075
[arXiv:1002.2085 [astro-ph.CO]].
  %%CITATION = ARXIV:1002.2085;%%


\bibitem{virgo2}
  R.~B.~Tully and E.~J.~Shaya,
  Astrophys.\  J.\  {\bf 281} (1984) 31.

\bibitem{virgo3}
 J.~L.~Tonry,  J.~P.~Blakeslee, E.~A.~Ajhar and A.~Dressler, 
  Astrophys.\  J.\  {\bf 530} (2000) 625.

\bibitem{virgo}
  P.~Fouque, J.~M.~Solanes, T.~Sanchis and C.~Balkowski,
  %``Structure, mass and distance of the virgo cluster from a Tolman-Bondi model,''
 Astronomy and Astrophysics {\bf 375 } (2001) 770
  [astro-ph/0106261].
  %%CITATION = ASTRO-PH/0106261;%%


\bibitem{fornax}
  O.~G.~Nasonova, J.~A.~d.~F.~Pacheco and I.~D.~Karachentsev,
  %``Hubble flow around Fornax cluster of galaxies,''
 Astronomy and Astrophysics {\bf 532} (2011) A104
 [arXiv:1106.1291 [astro-ph.CO]].
  %%CITATION = ARXIV:1106.1291;%%

\bibitem{makarov}
  D.~Makarov and I.~Karachentsev,
  %``Galaxy groups and clouds in the Local (z~0.01) universe,''
  Mon.\ Not.\ Roy.\ Astron.\ Soc.\  {\bf 412} (2011) 2498
  [arXiv:1011.6277 [astro-ph.CO]].
  %%CITATION = ARXIV:1011.6277;%%

\bibitem{eridanus}
  S.~Brough, D.~Forbes, V.~Kilborn, W.~Couch and M.~Colless,
  %``Eridanus: a supergroup in the local universe?,''
  Mon.\ Not.\ Roy.\ Astron.\ Soc.\  {\bf 369} (2006) 1351
  [astro-ph/0603778].
  %%CITATION = ASTRO-PH/0603778;%%

\bibitem{brouz}
  N.~Brouzakis and N.~Tetradis,
  %``Non-linear Matter Spectrum for a Variable Equation of State,''
  JCAP {\bf 1101} (2011) 024
  [arXiv:1002.3277 [astro-ph.CO]].
  %%CITATION = ARXIV:1002.3277;%%

\bibitem{ostriker} 
  K.~Subramanian, R.~Cen and J.~P.~Ostriker,
  %``The Structure of dark matter halos in hierarchical clustering theories,''
 Astrophys.\  J.\  {\bf 538} (2000) 528
  [astro-ph/9909279].
  %%CITATION = ASTRO-PH/9909279;%%

\bibitem{prada}
  H.~Tavio, A.~J.~Cuesta, F.~Prada, A.~A.~Klypin and M.~A.~Sanchez-Conde,
  %``The dark outside: the density profile of dark matter haloes beyond the virial radius,''
  arXiv:0807.3027 [astro-ph].
  %%CITATION = ARXIV:0807.3027;%%


\bibitem{kravtsov} 
 B.~Diemer and A.~V.~Kravtsov,
  %``Dependence of the outer density profiles of halos on their mass accretion rate,''
  arXiv:1401.1216 [astro-ph.CO].
 



\end{thebibliography}
\end{document}